\documentclass[aps,pre,twocolumn]{revtex4}
\usepackage{amsmath,bm,epsfig}

\usepackage{graphicx}
\usepackage{indentfirst}

\newcommand{\C}[1]{{\mathcal{#1}}}    
\newcommand{\sFrac}[2]{{\textstyle\frac{#1}{#2}}}
\begin{document}
\title{Scaling Theory for Steady State Plastic Flows in Amorphous Solids}
\author{Edan Lerner and Itamar Procaccia}
\affiliation{$^1$Department of Chemical Physics, The Weizmann Institute of Science, Rehovot 76100, Israel}
\
\date{\today}

\begin{abstract}
Strongly correlated amorphous solids are a class of glass-formers whose inter-particle potential admits an approximate inverse power-law form in a relevant range of
inter-particle distances. We study the steady-state plastic flow of such systems, firstly
in the athermal, quasi-static limit, and secondly at finite temperatures and strain rates. In all cases we demonstrate the
 usefulness of scaling concepts to reduce the data to universal scaling functions where the scaling exponents are determined a-priori from the inter-particle potential. In particular we show that the
steady plastic flow at finite temperatures with efficient heat extraction is uniquely characterized by two scaled variables;
  equivalently, the steady state displays an equation of state that relates one scaled variable to the other two. We discuss the range of applicability of the scaling theory, and the connection to density scaling in supercooled liquid dynamics. We explain that the description of transient states calls for additional state variables whose identity is still far from obvious.
\end{abstract}

\maketitle

\section{Introduction}

The equations of fluid mechanics appear to provide an adequate description for the flow of liquids for an extremely wide range of boundary conditions and external forcing. A similarly successful theory is still lacking for the description of
elasto-plastic dynamics in amorphous solids which form as the result of the glass transition. While being essentially
``frozen liquids", amorphous solids differ from regular liquids in having a yield strength $\sigma_s$, a material parameter which depends on the density, temperature etc, which is the maximal value of the
internal stress that the material can support by elastic forces. Regular liquids cannot support any amount of stress without flowing. When the stress exceeds the yield strength the material begins to respond plastically, and under a given external shear
rate can develop a steady state plastic flow with a mean ``flow stress" $\sigma_\infty$. The analog of the Navier-Stokes
equations which can describe the whole spectrum of elasto-plastic responses in terms of macroscopic variables is not
known yet, and their derivation is the subject of much current research \cite{98FL,05DA,09LPCH,04ML,06TLB,06ML,07BSLJ,09LP,08TTLB}
with significant amount of debate. In this paper we
focus attention on the steady-state plastic flow which is obtained under the action of a constant external strain rate. We will argue below that the characterization of such a state is considerably simpler than the full description of transient states, the latter call for a larger number of macroscopic variables whose nature is not obvious and the constitutive relations between them are not known. For the steady plastic flow state we can make progress and determine what are the state variables that determine
the state uniquely.

To simplify things further we limit our attention at present to materials whose inter-particle potential can be approximated,
for the range of inter-particle distances of relevance, by an inverse power law potential.  This same class of materials and the interesting scaling properties that they exhibit attracted considerable interest in the context of the dynamics of super-cooled liquids, first experimentally \cite{04CR,04SCAT,04DGGSP,05CR} and then theoretically \cite{08IPRS,08PBSD,08CR,08BPGSD}.  In the context of the mechanical properties of amorphous solids we believe that the first example of using the special scaling properties of these materials appeared in \cite{09LPCH} where focus was put on the athermal limit and quasi-static strain.
In this paper we explore further the quasi-static limit, and then extend the discussion to systems at finite temperatures
and finite strain rates. The discussion culminates with finding which are the minimal number of re-scaled state variables
that determine uniquely the steady plastic flow in such materials. Any general theory that attempts to provide a complete
description of elasto-plasticity in amorphous solids should reduce, in the steady flow state of materials of the present class, to a theory that contains these and only these variables.

The structure of the paper is as follows: In Sect. \ref{system} we introduce the systems under study, and
explain how they are simulated both in the athermal, quasi-static limit and at finite temperatures and strain rates.
In Sect. \ref{scaling} we explain the special scaling properties that these systems possess, and predict theoretically
what is expected in the steady plastic flow state. This is the central part of the paper. We then
provide detailed presentations of simulation results and demonstrate how they compare to the predictions of the scaling theory.
We discuss analytic properties of the scaling function, and demonstrate the conditions under which the scaling breaks down.
In Sect. \ref{supercooled} we discuss the consequences of our thinking to supercooled liquids, and propose that the
scaling function used in the literature in this context are incomplete.
Sect. \ref{summary} summarizes the findings, and provides a discussion of the road ahead,
especially in terms of extensions to transient states.

\section{Systems and Methods of Simulation}
\label{system}
\subsection{System Definitions}
In this work we employ two-dimensional polydisperse systems of
point particles of equal mass $m$,
interacting via two qualitatively different pair-wise potentials.
Each particle $i$ is assigned an
interaction parameter $\lambda_i$ from a normal distribution with mean $\langle \lambda_i \rangle = 1$.
The variance is governed by the poly-dispersity parameter $\Delta = 15\%$ where $\Delta^2 = \frac{\langle \left(\lambda_i -
\langle \lambda \rangle \right)^2\rangle}{\langle \lambda \rangle^2} $.
With the definition $\lambda_{ij} = \textstyle\frac{1}{2}(\lambda_i + \lambda_j)$
the first potential $U_R(r_{ij})$ is purely repulsive, of which the shape is
characterized by the interger $k$:
\begin{widetext}
\begin{equation}
U_R(r_{ij}) =
\left\{
\begin{array}{ccl}
\!\!\! \epsilon\left[\left(\frac{\lambda_{ij}}{r_{ij}}\right)^{k}\!\! -\!\!\frac{k(k+2)}{8}
\left( \frac{B_0}{k} \right)^{\frac{k+4}{k+2}}\left(\frac{r_{ij}}{\lambda_{ij}}\right)^4
+ \frac{B_0(k+4)}{4}\left(\frac{r_{ij}}{\lambda_{ij}}\right)^2
-\frac{(k+2)(k+4)}{8}\left( \frac{B_0}{k} \right)^{\frac{k}{k+2}}\right] & , & r_{ij} \le
\lambda_{ij}\left( \frac{k}{B_0} \right)^{\frac{1}{k+2}} \\
0 & , & r_{ij} >
\lambda_{ij}\left( \frac{k}{B_0} \right)^{\frac{1}{k+2}}
\label{potential}
\end{array}
\right\}\ ,
\end{equation}
\end{widetext}
We chose $B_0 = 0.2$ for all systems discussed, and vary the integer $k$ in the following.
This pair-wise potential
is constructed such as to minimize computation time, and is smooth up to second
derivative, which is required for minimization procedures.

\begin{figure}[ht]
\centering
\includegraphics[scale = 0.45]{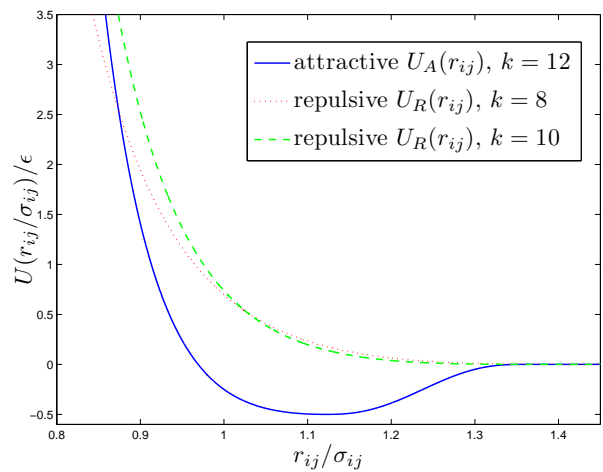}
\caption{\footnotesize Color online: The different pairwise potentials discussed in this work.}
\label{potentialsFig}
\end{figure}

The second pair-wise potential $U_A(r_{ij})$ reads
\begin{equation}
U_A(r_{ij}) = \left\{
\begin{array}{ccl}
\tilde{U}(r_{ij})  & , & r \le r_\star(\lambda_{ij}) \\
\hat{U}(r_{ij})  & , & r_\star(\lambda_{ij}) < r  \le r_c(\lambda_{ij}) \\
0 & , & r > r_c(\lambda_{ij})
\end{array}
\right.
\label{newpot}
\end{equation}
with $\tilde{U}(r_{ij}) = \epsilon\left[\left(\frac{\lambda_{ij}}{r_{ij}}\right)^k -
\left(\frac{\lambda_{ij}}{r_{ij}}\right)^6 - 1/4 \right]$; $k=12$, $r_\star = 2^{1/6} \lambda_{ij}$ and $r_c=1.36\lambda_{ij}$.
The attractive part $\hat U(r)$ is glued smoothly to the repulsive part.
We choose  $\hat U(r) =\frac{\epsilon}{2} P\left(\frac{r-r_0}{r_c-r_0}\right)$ where $P(x) = \sum_{i=0}^5A_ix^i$  and the
coefficients $A_i$ (see Table~\ref{table}) are chosen such that the potential is smooth up to second derivative.
\begin{table}[ht]
\begin{tabular}{|c|c|}
\hline
$A_0$&-1.0\\
$A_1$&0.0\\
$A_2$&0.806111631332424\\
$A_3$&7.581665106002721\\
$A_4$&-12.581665106002717\\
$A_5$&5.193888368667571\\
\hline
\end{tabular}
\caption{The coefficients in $P(x) = \sum_{i=0}^5A_ix^i$, see text.}
\label{table}
\end{table}
These pairwise potentials are displayed in Fig.~\ref{potentialsFig} for the cases of interest.
Below the units of length, energy, mass and temperature are $\lambda \equiv \langle \lambda_i \rangle$, $\epsilon$, $m$ and $\epsilon/k_B$ where
$k_B$ is Boltzmann's constant.
The time units $\tau_\star$ are accordingly $\tau_\star=\sqrt{m \lambda^2/\epsilon}$.
From here and in the following
we denote the density as $\tilde{\rho} \equiv \frac{N}{V}$, and define the dimensionless density
$\rho \equiv \lambda^2\tilde{\rho}$. Also, we will refer to the dimensionless density
as just the density, for the sake of brevity.

Initial conditions for all the simulations, for both methods described in the next Subsection,
were obtained by instantaneous quenching
of random, high temperature configurations; this explains the apparent noise and
absence of stress peaks in the transients. Furthermore, it is important to note that
due to finite system sizes, the initial value
of the stress of the quenched configurations in some experiments
is non-zero; this is however irrelevant for steady state statistics.

\subsection{methods}
The work presented here is based on two types of simulational methods. The first type
corresponds to the athermal quasi-static (AQS)
limit $T \rightarrow 0$ and $\dot{\gamma} \rightarrow 0$,
where $\dot{\gamma}$ is the strain rate.
AQS methods have been extensively used
recently \cite{04ML,06TLB,06ML,07BSLJ,08TTLB,09LP} as a tool for investigating
plasticity in amorphous systems.
The order in which the limits $T \rightarrow 0, \dot{\gamma} \rightarrow 0$
are taken is important, since one expects that at
any finite temperature the stress in the system can thermally relax
given long enough time \cite{08EP} (or small enough strain rates),
hence the limit $T \rightarrow 0$ should be taken
prior to the $\dot{\gamma} \rightarrow 0$ limit.
According to AQS methods, starting from
a completely quenched configuration of the system, we apply an affine simple shear
transformation to each particle $i$ in our shear cell, according to
\begin{eqnarray}\label{affineTransformation}
r_{ix} & \rightarrow & r_{ix} + r_{iy}\delta\epsilon\ , \nonumber\\
r_{iy} & \rightarrow & r_{iy} \ ,
\label{simpleShearTransformation}
\end{eqnarray}
in addition to imposing Lees-Edwards boundary conditions \cite{91AT}.
The strain increment $\delta\epsilon$ plays a role analogous to the integration
step in standard MD simulations. We choose for the discussed systems $\delta\epsilon = 10^{-4}$,
which while not sufficiently small for extracting exact statistics of plastic flow events as done in
\cite{09LP}, it is, however,
sufficiently small for the analysis of the steady state properties and mean values.
The affine transformation (\ref{affineTransformation})
is then followed by the minimization \cite{minimizer} of the potential
energy under the constraints imposed by the strain increment and the periodic
boundary conditions. We chose
the termination threshold of the minimizations to be $|\nabla U|^2/N = 10^{-18}$.

The second simulation method employs the so-called SLLOD equations of motion
\cite{91AT}. For our constant strain rate 2D systems, they read
\begin{eqnarray}
\dot{r}_{ix} & = & p_{ix}/m + \dot{\gamma}r_{iy}\ ,\nonumber \\
\dot{r}_{iy} & = & p_{iy}/m\ , \nonumber \\
\dot{p}_{ix} & = & f_{ix} - \dot{\gamma}p_{iy}\ , \nonumber \\
\dot{p}_{iy} & = & f_{iy}\ . \nonumber
\end{eqnarray}
We use a leapfrog integration scheme for the above equations,
and keep the temperature constant by employing the Berendsen thermostat \cite{91AT},
measuring the instantaneous temperature with respect to a homogeneous shear flow.
The integration time steps were varied between $\delta t = 0.007$ and
$\delta t = 0.001$, depending on density, such that numerical stability
was maintained for all densities simulated.
The time scale $\tau_T$ for heat extraction \cite{91AT} was chosen such that rate of heat generation is smaller
than the rate of heat extraction. For the lowest densities this was chosen to be $\tau_T \approx
10\tau_\star$.

\section{The Scaling Theory}
\label{scaling}

\begin{figure}
\centering
\includegraphics[scale = 0.43]{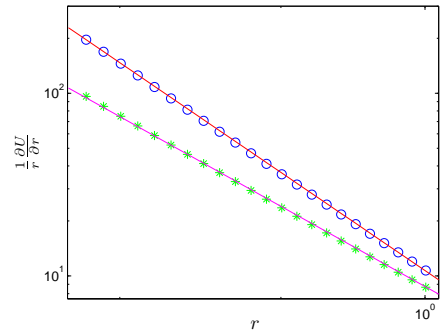}
\caption{Color online: $r^{-1} \frac{\partial U(r)}{\partial r}$ in the range of $r_0/\lambda \in [\rho_{\rm max}^{-1/2},\rho_{\rm min}^{-1/2}]$
for $k=8$ in green asterisks, and for $k=10$ in blue circles.
The line through the points represents the scaling laws (\ref{fundsc}).}
\label{pairwisePotentialScaling}
\end{figure}

The discussion of the relaxation properties of glass formers in the super-cooled regime \cite{04CR,04SCAT,04DGGSP,05CR,08IPRS,08PBSD,08CR,08BPGSD} and of the mechanical properties of the amorphous solids \cite{09LPCH} simplifies significantly when the inter-particle potential assumes
an effective inverse power-law from in the relevant range of inter-particle distances.
As an example consider the potential (\ref{potential}) in the density range $\rho\in [1,1.6]$. Since in $d$ dimensions the characteristic inter-particle distance $r_0$ scales like
\begin{equation}
r_0\sim \frac{\lambda}{\rho^{1/d}} \ ,
\label{r0}
\end{equation}
the range of densities employed here is equivalent to a range of $r_0/\lambda\in [\rho_{\rm max}^{-1/d},\rho_{\rm min}^{-1/d}]$. We find that in this range, to a very good approximation,
 \begin{equation}
\frac{1}{r^{d-1}}\frac{\partial U_R(r)}{\partial r} \sim \frac{\epsilon}{\lambda^d} \left(\frac{r}{\lambda}\right)^{-\nu d} \ .
\label{fundsc}
\end{equation}
In two dimensions $\nu=4.80$ for $k=8$ and $\nu=5.87$ for $k=10$, see Fig.~\ref{pairwisePotentialScaling}.

In the following discussion we define the flow stress $\sigma_\infty$ to be the steady-state
value of the stress under constant external strain rate. In general,
the flow stress is a function of a set of state variables, which specify
the conditions in which the experiments are carried out.
For the systems and experiments discussed in this work, the flow stress
depends on the density $\rho$,
the temperature $T$, and the strain rate $\dot{\gamma}$. In addition, one can expect also a
dependence on the heat extraction rate $\tau_T^{-1}$. We choose to exclude the latter from the present
discussion, and we do so by choosing the rate of heat extraction to be much larger than the
rate of heat production. So, we propose at this
point that $\sigma_\infty = \sigma_\infty(T,\rho,\dot\gamma)$.
The yield stress $\sigma_Y(\rho)$ is defined as the steady state value of the stress
under the limits $T \rightarrow 0$ and $\dot{\gamma} \rightarrow 0$ (see discussion regarding
these limits in Subsect.~\ref{analytic}), i.e.
\begin{equation}\label{yieldStressDefinition}
\sigma_Y \equiv \sigma_\infty(\rho, T \rightarrow 0, \dot{\gamma} \rightarrow 0)\ .
\end{equation}
\subsection{Scaling in the Athermal, Quasi-static limit}

In the athermal, quasi-static limit the only parameter left is the density;
consideration of the temperature and strain rate effects will be taken up in the next Subsection.
Denote the distribution of inter-particle distances as $p(r)$; then
the mean inter-particle distance is $r_0(\rho) \equiv \int rp(r;\rho)dr$.
Note that this probability distribution only accounts for distances which
are relevant in terms of the interaction, namely for $r_{ij} \le
\lambda_{ij}\left( \frac{k}{B_0} \right)^{\frac{1}{k+2}}$.
If $p(r)$ is sufficiently sharply peaked around $r_0$, we can write
\begin{equation}\label{moo2}
\left< r \frac{\partial U_R}{\partial r} \right> \sim
r_0 \left. \frac{\partial U_R}{\partial r}\right|_{r_0} \sim
\epsilon\left(\frac{r_0}{\lambda}\right)^{d(1-\nu)}\sim \epsilon\rho^{\nu -1}\ .
\end{equation}
From here we predict that for our systems with short-range forces the scaling of the yield stress should be
\begin{equation}\label{ysscaling}
\sigma_Y \sim N\frac{r_0 \left. \frac{\partial U_R}{\partial r}\right|_{r_0}}{V} \sim \sFrac{\epsilon}{\lambda^d}\rho^\nu\ .
\end{equation}
In the athermal, quasi-static limit the shear modulus must obey the same scaling
\begin{equation}\label{smscaling}
\mu \sim \sFrac{\epsilon}{\lambda^d}\rho^\nu\ .
\end{equation}
\begin{figure}
\centering
\includegraphics[scale = 0.45]{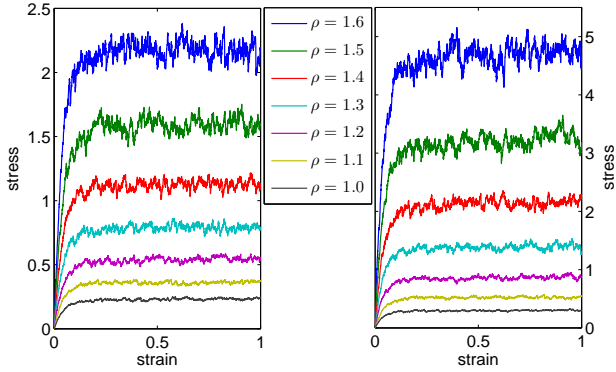}
\caption{Color online: stress-strain curves averaged over 20 independent runs for an athermal system with $N=4096$, $k=8$ (left panel) and $k=10$ (right panel) as a function of the density, with the density increasing from bottom to top.}
\label{rawsig}
\end{figure}
\begin{figure}
\centering
\includegraphics[scale = 0.39]{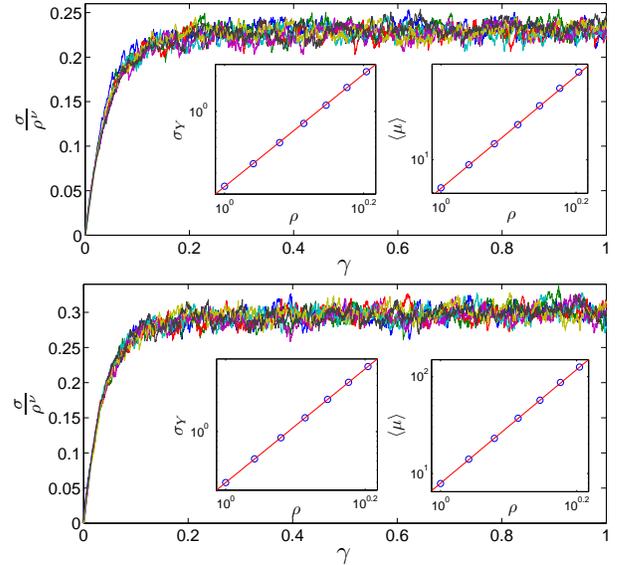}
\caption{Color online: The same stress-strain curves as in Fig.~\ref{rawsig} but with the stress rescaled by $\rho^\nu$, with
$\nu=4.80$ for $k=8$ (top panel) and $\nu=5.87$ for $k=10$ (bottom panel). The insets demonstrate the density dependence of $\sigma_Y$ and $\mu$ according
to $\rho^\nu$. }
\label{scalesig}
\end{figure}
These scaling laws lead to the expectation that re-plotting stress-strain curves in terms of re-scaled variable $\sigma/\rho^\nu$ should result in complete data collapse. Indeed, our simulations vindicate this expectation. In Fig.~\ref{rawsig} we present the raw stress-strain curves in the athermal, quasi-static limit using seven different values
of the density. For each density we simulated 20 independent runs of $N = 4096$ particles, using the pairwise potential (\ref{potential}) and two choices of the integers $k=8$ and $k=10$. Fig.~\ref{scalesig} demonstrates the superb data collapse for the scaled variable. The insets are a direct test
of the scaling laws (\ref{ysscaling}) and (\ref{smscaling}).

\subsection{Scaling Theory with Temperature and External Strain Rate}
\label{Tscaling}
Once we perform measurements at finite temperatures and external strain rates the scaling considerations must
incorporate temporal and energy scales. The typical free energy density in the steady-state plastic flow should scale
like $\sigma_Y \times \delta \epsilon$ where $\delta \epsilon$ is the typical strain interval between plastic events,
$\delta \epsilon \sim \sigma_Y/\mu$. Accordingly, the intensive energetic contribution to barriers $\delta G$
(that govern thermal activation) scales with the density according to
\begin{equation}
\langle \delta G \rangle \sim \frac{V\frac{\sigma_Y^2}{\mu}}{N} \sim \epsilon\rho^{\nu-1}\ .
\label{deltaG}
\end{equation}
Note that this is the ``density scaling" proposed in \cite{04CR,04SCAT,04DGGSP,05CR,08IPRS,08PBSD,08CR,08BPGSD}
in the context of the dynamics of super-cooled liquids.
For the present purposes we need to explore further scaling relations;
we estimate now the density scaling of the typical time-scale $\tau_0$
with respect to which all the rates in the theory should be compared. We begin with the speed of sound $c_s$; using
Eq.~(\ref{smscaling}) we write
\begin{equation}
c_s = \sqrt{\frac{\mu}{\rho}} \sim \sFrac{\lambda}{\tau_\star}\rho^{\frac{\nu-1}{2}}\ .
\label{cs}
\end{equation}
We can now define the time scale $\tau_0\equiv r_0/c_s$; Using Eqs.~(\ref{r0}) and (\ref{cs}) we obtain
\begin{equation}
\tau_0 \sim \tau_\star\rho^{-\frac{\nu d - d + 2}{2d}}\ .
\label{tau0}
\end{equation}
Using Eq. \ref{deltaG} we conclude that the effect of temperature on the dynamics in the steady state
must be invariant once the temperature is rescaled by $\rho^{\nu-1}$. On the other hand the external
strain rate $\dot \gamma$ should leave the system invariant once rescaled by $\rho^{-\frac{\nu d - d + 2}{2d}}$ due to Eq.~\ref{tau0}.
Putting together all these
we finally propose the expected scaling-function form for the flow stress $\sigma_{\infty}$:
\begin{equation}\label{moo1}
\boxed{\sigma_{\infty}( T, \rho, \dot{\gamma} ) = \sFrac{\epsilon}{\lambda^d}\rho^\nu {\cal S}
\left(\frac{T}{\epsilon\rho^{\nu-1}} , \frac{\dot{\gamma}}{\tau_\star^{-1}\rho^{\frac{\nu d - d + 2}{2d}}} \right)}\ .
\end{equation}
This is the central theoretical result of this section.
We stress that we chose to favor the flow stress and wrote it
in terms of the scaling function of the other two dimensionless variables. We could equivalently choose any of the other
two variables to be represented in an analog way in terms of two dimensionless variables. This scaling function form
is in fact an equation of state for the steady plastic flow.

\begin{figure}[ht]
\centering
\includegraphics[scale = 0.48]{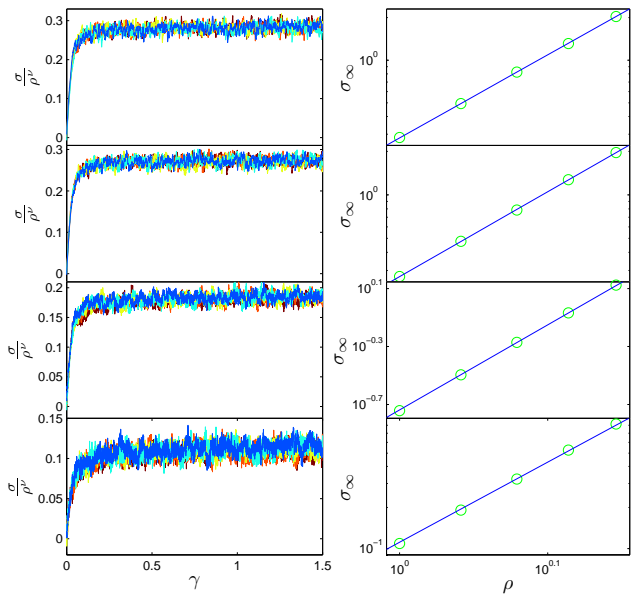}
\caption{Color online: Left Panels: stress normalized by $\rho^\nu$ vs. strain for the $x$ values
$x = 0.001, x = 0.01, x = 0.1$ and $x = 0.2$, increasing from top to bottom.
Right panels: log-log plots of the steady state flow stress as a function of density, for the same
corresponding values of $x$.}
\label{all_x}
\end{figure}

For $d=2$ this general result assumes the form
\begin{equation}\label{moo1}
\sigma_{\infty}( T, \rho, \dot{\gamma} ) = \sFrac{\epsilon}{\lambda^2}\rho^\nu {\cal S}
\left(\frac{T}{\epsilon\rho^{\nu-1}} , \frac{\dot{\gamma}}{\tau_\star^{-1}\rho^{\nu/2}} \right)\ .
\end{equation}
To demonstrate the high degree of precision with which the scaling theory is obeyed we performed simulations
at finite temperature and strain rate (see methods section) in which
we prepared 10 independent systems (for each density) of $N=10000$ particles
at the densities $\rho = 1.0,1.1,1.2,1.3$ and 1.4. Defining
the two dimensionless variable $x \equiv \frac{T}{\epsilon\rho^{\nu-1}}$ and $y \equiv \frac{\dot{\gamma}}{\tau_\star^{-1}\rho^{\nu/2}}$, we fix the value $y_0 = 1.6\times10^{-5}$
for all densities, and simulated all the five densities for the values $x = 0.001, 0.01,
0.1$ and 0.2. The results are displayed in Fig. ~(\ref{all_x}).
We see the excellent data collapse and also the quality of the scaling laws for the flow stress; the slopes of the lines in
the right panels are those predicted theoretically in Eq. (\ref{moo1}), i.e. $\sigma_\infty \sim \frac{\epsilon}{\lambda^2}\rho^\nu$.

\begin{figure}[ht]
\centering
\includegraphics[scale = 0.52]{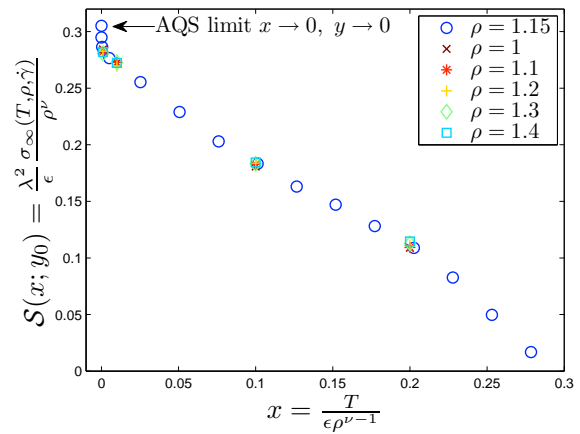}
\caption{Color online: The function ${\cal S}(x;y_0)$. Data is displayed for $\rho = 1.15$ (blue circles)
over a wide range of
$x = \frac{T}{\epsilon\rho^{\nu-1}}$ values, and for the densities of
Fig.~\ref{all_x} over the $x$ values $x = 0.001, x = 0.01, x = 0.1$ and $x = 0.2$. The value of $y_0 = \frac{\dot{\gamma}}{\tau_\star^{-1}\rho^{\nu/2}}$
is $1.658\times10^{-5}$
for all simulated systems.}
\label{sOFx}
\end{figure}

We now test the quality of the prediction of the existence of the scaling function $\C S(x,y)$. To this
aim we fixed a value of $\rho=1.15$ and the same $y_0 = 1.6\times10^{-5}$, and simulated the entire range
of $x$ values for which $\C S(x,y)$ exists. The result is shown in Fig. \ref{sOFx}, in addition to the data
obtained for all the other densities and $x$ values shown in Fig. \ref{all_x}. The excellent data collapse is quite apparent.
It is noteworthy that at low temperatures the function reaches smoothly, albeit with a very high gradient,
precisely the athermal, quasi-static limit that was studied in the previous Subsection. The high gradient as $T\to 0$
in a similar, experimentally obtained function, was interpreted in \cite{05JS} as resulting from quantum-mechanical effects.
Obviously in our purely classical simulations there are no quantum effects and it remains very interesting to unfathom
the origin of the very fast change in the flow stress over a very short temperature interval.

\begin{figure}[ht]
\centering
\includegraphics[scale = 0.55]{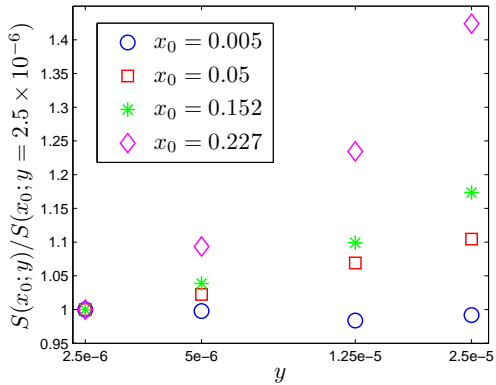}
\caption{Color online: The scaling function ${\cal S}(x_0,y)$ normalized by the values
${\cal S}(x_0,y = 2.5\times 10^{-6})$, for various values of $y$.}
\label{strainRateDependence}
\end{figure}

To emphasize the relevance of the temporal scaling we simulated steady flow states at different external
strain rates but at the same $x$ values. The result are shown in Fig. \ref{strainRateDependence}.
We see that as the temperature increases, the relative sensitivity of the flow stress to changes in the
the strain rate increases appreciably. Note that the value of $y_0 = 1.6\times10^{-5}$ for which the data collapse was demonstrated is well within the range of high sensitivity to changes in the strain rate. In other words, without rescaling the
strain rate properly there is no hope for data collapse. Further analytic properties of the scaling function
are discussed in the next Subsection.
\subsection{Analytic Properties of the Scaling Function}
\label{analytic}

The entire physics of the steady flow state for this class of systems is encoded in the scaling function $\C S(x,y)$.
It is therefore very challenging to derive the form of this functions from first principles. We are not yet in a position
to do so; at this point we can only present the analytic properties of this function as a preparation for future
discussions.

Firstly, it is noteworthy that the limits $\lim_{x\to 0}\lim_{y\to 0}$ and $\lim_{y\to 0}\lim_{x\to 0}$ do not commute.
We expect that
\begin{equation}
\lim_{x\to 0}\lim_{y\to 0} \C S(x,y) =0 \ ,
\end{equation}
simply because at any finite temperature, given enough time to relax the stress, the flow stress must vanish \cite{08EP}.
On the other hand
\begin{equation}
\lim_{y\to 0}\lim_{x\to 0} \C S(x,y) =\sigma_Y/\rho^\nu \ ,
\end{equation}
as can be seen directly from Fig. \ref{sOFx}.

Secondly, in the athermal limit $x\to 0$ the flow stress loses its dependence on the external strain rate for
sufficiently small values of $y$,
\begin{equation}
\lim_{y\to 0}\lim_{x\to 0} \frac{\partial\C S(x,y)}{\partial y} =0 \ .
\end{equation}
This property can be seen directly in Fig. \ref{strainRateDependence}. The physical reason for this property is
that without substantial thermal activation the physics becomes insensitive to external time scales. This limit is expected to
hold when the external strain rate is much smaller than the elastic relaxation rate; interplays between high
strain rates and the flow stress were investigated in \cite{09LC}.

Finally, we observe an inflection point in $\C S(x,y)$, see Fig. \ref{sOFx}, where
\begin{equation}\label{inflectionPoint}
\left. \frac{\partial^2 \C S(x,y)}{\partial x^2}\right|_y =0 \ .
\end{equation}
We conjecture that this inflection point separates a ``low temperature region" from a ``high temperature region"
in which the elasto-plastic physics is not the same. It is possible that a change from delocalized plastic events
to more localized events \cite{09LC,09LP} is the fundamental reason for this change, but further study is necessary to pinpoint
this issue in a convincing way.

\subsection{Applicability of the Scaling Theory}

At this point it is appropriate to discuss the general applicability of the scaling approach. It is sufficient
to delineate this applicability in the context of the athermal, quasi-static limit using systems in which the inter-particle potential cannot be usefully approximated as inverse power laws. In some model systems, e.g. \cite{08CR}, it has been
shown that density scaling of the dynamics of super-cooled liquids still holds in spite of the presence of
attractive forces in the  potential. Furthermore, the same qualitative density scaling
has been applied to a wide variety of experimental data, with substantial success \cite{04CR,04SCAT,04DGGSP,05CR}.
In these experimental systems there are definitely attractive forces between the particles, and thus the question of
the applicability of the scaling theory is highly pertinent.

\begin{figure}
\centering
\includegraphics[scale = 0.43]{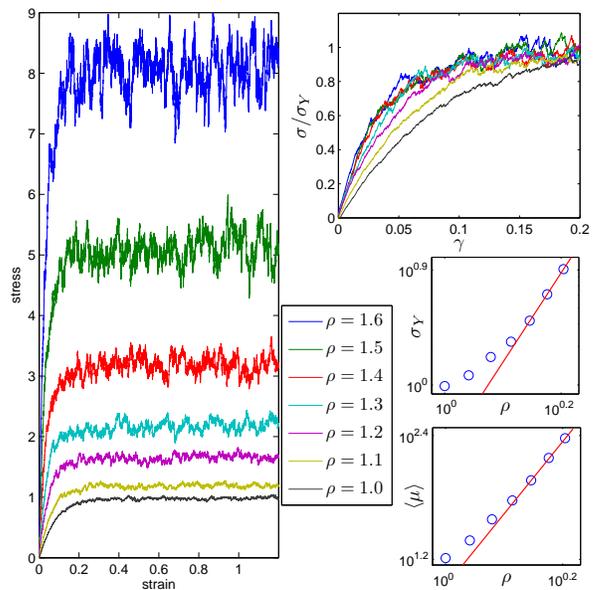}
\caption{Color online. Left panel: stress-strain curves for the potential (\ref{newpot}) which has a repulsive and an attractive part. Right upper panel: demonstration of the failure of rescaling of the stress-strain curves. Lower panels: $\sigma_Y$ and $\langle \mu\rangle $ as a function of the density.  Note that predictability is regained only for higher densities, the straight line is $\rho^7$.}
\label{noScaling}
\end{figure}

\subsubsection{Simulations}
We have simulated systems with the potential $U_A(r)$, Eq.~(\ref{newpot}) in the athermal, quasi-static limit. In this potential an attractive branch is added to the repulsive one, see Fig.~\ref{potentialsFig}.
We again prepared 20 independent runs for each of the 7 densities $\rho = 1.0, 1.1, 1.2, 1.3, 1.4, 1.5$ and 1.6,
this time for systems of $N=2500$ particles, and collected statistics for the steady state stress values (see methods section),
as previously described.

The raw data of the stress-strain curves is displayed in the left panel of Fig.~\ref{noScaling}.
In the right upper panel we show what happens when we try to collapse the data by rescaling
the stress by $\sigma_Y$. Of course the stress-stain curves now all asymptote to the same value, but the curves fail to collapse,
since $\langle \mu\rangle $ does not scale in the same way as $\sigma_Y$. Nevertheless, even in the present case we can have predictive power for high densities. When the density increases the repulsive part of the
potential (\ref{newpot}) becomes increasingly more relevant, and the inner power law $r^{-12}$ becomes dominant. We therefore
expect that for higher densities scaling will be regained, and both $\sigma_Y$ and $\langle \mu \rangle$ would depend on the density as $\rho^7$.
The two lower right panels in Fig.~\ref{noScaling} show how well this prediction is realized also in the present case.

\subsubsection{Constancy of the ratio of the shear modulus and the yield stress}
Another way of flushing out the failure of scaling when there exist attractive forces is provided by the ratio
\begin{equation}\label{omegaDefinition}
\Omega \equiv \frac{\mu}{\sigma_Y}\ .
\end{equation}
This is a pure number, which has been claimed to be universal for a family of metallic glasses \cite{05JS}.
For systems in which our scaling analysis holds, we have seen that the shear modulus scales with density in
exactly the same manner as the yield stress (see Eq.~(\ref{ysscaling}),(\ref{smscaling})), hence the number $\Omega$ should be invariant to density changes, for a given system. However, when compared across different systems, there is no a-priori
reason to expect this number to be universal.
\begin{figure}
\centering
\includegraphics[scale = 0.48]{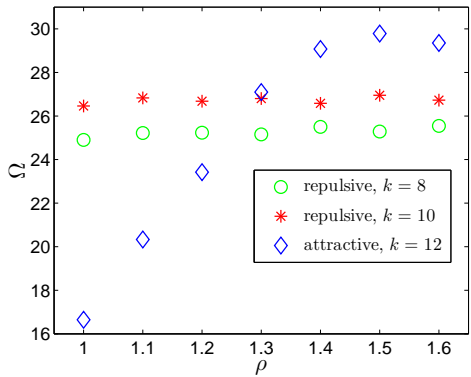}
\caption{Color online: The pure number $\Omega$ as a function of the density for the three potentials discussed in the text. Note that $\Omega$
appears to increase with the exponent of the repulsive part of the potential whenever scaling prevails.}
\label{omegaFig}
\end{figure}
Fig.~\ref{omegaFig} displays the measured values of $\Omega$ for our athermal, quasi-static
experiments, for two different repulsive potentials of the form (\ref{potential}), using $k=8$ and $k=10$, and for the
attractive potential (\ref{newpot}), with $k=12$.
For the two repulsive potentials, we find from our numerics that
this parameter differs by about 5\%, indicating non-universality. The lack of universality is even clearer
with the last potential (\ref{newpot}).
It is apparent that when scaling prevails the value of $\Omega$
is constant up to numerical fluctuations. In the third case, where scaling fails, $\Omega$ is a strong function of $\rho$ except
at higher densities where scaling behavior is recaptured as explained. We can therefore conclude that the approximate constancy
of $\Omega$ found in a family of metallic glasses \cite{05JS}, is not fundamental but only an indication of the similarity
of the potentials for this family. In general $\Omega$ can depend on the inter-particle potential.
It is quite clear from considering Eqs.~(\ref{moo2}), (\ref{ysscaling}) and (\ref{smscaling}),
that the {\em coefficients} in the scaling laws (\ref{ysscaling}) and (\ref{smscaling})
may well depend on the exponent $k$ in the
repulsive part of the potential. The ratio of these pre-factors, being a pure number, could be
independent of $k$, and $\Omega$ could be universal. It appears however that $\langle \mu \rangle$ is increasing more
with $k$ than $\sigma_Y$, and therefore $\Omega$ shows a clear increase upon increasing $k$.
At present this must remain an interesting riddle for future research.

\section{Relation to density scaling in supercooled liquids}
\label{supercooled}

The destruction of scaling for low-density systems with the attractive potential (\ref{newpot}) is in apparent
contradiction to density scaling analysis of relaxation times in supercooled liquids. As mentioned above, it has been shown
in the context of the dynamics of supercooled liquids, both in model systems and in experiments,
that the presence of attractive forces in the pairwise potentials can still be consistent with density
scaling. In our context of mechanical properties scaling is regained only at high densities; it is
desirable to understand whether there is a qualitative difference between the influence of attractive forces
on mechanical properties, and the influence of attractive forces on the dynamics of supercooled liquids.

The standard way in which density scaling is presented in the context of the dynamics of supercooled liquids
is in the form \cite{04CR,04SCAT,04DGGSP,05CR,08PBSD,08CR}
\begin{equation}\label{supercooledScaling}
\tau_\alpha(T,\rho) = {\cal F}\left(\frac{T}{\rho^{\gamma}}\right)\ ,
\end{equation}
where $\tau_\alpha$ is the $\alpha$-relaxation time and ${\cal F}(x)$ is a scaling function of one rescaled variable;
the exponent $\gamma$ corresponds to $\nu-1$ in our scaling analysis.

In our opinion this form cannot be exact, and we propose now an alternative form in light of the analysis
presented above. The form (\ref{supercooledScaling}) account only for the density scaling of the free-energy barriers for thermal activation. We have noted above that on top of this the microscopic time-scale $\tau_0$, with respect to which rates are compared, also varies with density, see Eq. (\ref{tau0}) and discussion in Subsect.~\ref{Tscaling}.

Write the $\alpha$-relaxation time in the standard transition-state-theory form
\begin{equation}\label{TST}
\tau_\alpha(T,\rho) = \tau_0 e^{\frac{\delta G(T)}{T}}\ .
\end{equation}
The free-energy barrier $\delta G$ scales with density as $\delta G \sim \epsilon\rho^{\nu-1}$ (see discussion
prior to Eq.~(\ref{deltaG}) ); the microscopic time scale should scale as $\tau_0 \sim \tau_\star\rho^{-\frac{\nu d - d + 2}{2d}}$, (see discussion prior to Eq.~(\ref{tau0}) ).
Combining these considerations, we obtain the scaling form
\begin{equation}\label{proposedSupercooledScaling1}
\tau_\alpha(T,\rho) = \tau_\star\rho^{-\frac{\nu d - d + 2}{2d}}{\cal F}\left(\frac{T}{\epsilon\rho^{\nu-1}}\right)\ .
\end{equation}

We believe that this correct form was missed because the
scaling of thermal activation barriers appears in the exponent of the RHS of (\ref{TST}), whereas the
scaling of the microscopic time scale is in the pre-factor.  Nevertheless it is our suggestion that
data should be re-analyzed using the proper form of the scaling function.

\section{Summary and the Road Ahead}
\label{summary}

In this paper we offered some modest inroads into providing a theory for elasto-plastic dynamics.
We must admit that
a complete theory of elasto-plastic response of amorphous solids is still out of reach, mainly because
of some fundamental riddles that are highly debated. Our proposition in this paper is that understanding
the steady plastic flow state is firstly simpler than and secondly mandatory for achieving a full theory
of elasto-plasticity. By focusing on glass formers with simple effective inverse power-law potentials we achieved
a scaling theory for the steady-state flow stress under constant
strain rate and finite temperatures. We have shown that in the athermal, quasi-static limit the yield stress
exhibits power-law dependence on the density, as does the shear modulus. It was then shown that
temperature and external strain rate can be incorporated into the scaling approach by accounting for
thermal activation effects via energy scaling, and rate effects via temporal scaling. The finite temperature
and finite strain rate theory appears in excellent agreement with the athermal, quasi-static limit when the
appropriate limits are taken.

The first task ahead is to provide an understanding from first principles of the scaling function ${\cal S}(x,y)$.
We have discussed some analytical properties of this scaling function, some of which
offer fascinating riddles for future research. Probably the most intriguing of these is the
inflection point in ${\cal S}(x,y)$, see Eq.~(\ref{inflectionPoint}) and the corresponding discussion.
Understanding the origin of this
inflection point may shed light on the possibility of constructing mean field theories of plasticity at least
for steady states, including the external parameter regimes for which they might be valid.

Probably the most important remaining issue is the identification of additional state-variable that
are necessary to describe transient states. It is well known that after straining in one direction and reaching
a steady state, a change in straining direction with an angle with respect to the original direction results in
angle dependent trajectories. This means that a tensorial order parameter is written into the material
during the steady flow state, and this object does not appear in our analysis. It must appear however
in the transient trajectories. The identification of this tensorial object will call for additional future work.

\acknowledgments
We have benefitted from e-mail discussions with George Hentschel. This work has been supported in part by
the Israel Science Foundation, the German Israeli Foundation and the Minerva Foundation, Munich, Germany.

\end{document}